# Giant second-harmonic generation enabled by bound-state continuum cavity on lithium niobate membrane

Juan José Robayo Yepes,[1,2] Fadi Issam Baida,[2] and Abdoulaye Ndao[1,*]

[1]Department of Electrical and Computer Engineering & Photonics Center, Boston University, 8 Saint 8 Mary's Street, Boston, Massachusetts 02215, USA
[2]Institut FEMTO-ST, UMR 6174 CNRS, Département d'Optique P. M. Duffieux, Université Bourgogne 6 Franche–Comté, 25030 Besançon Cedex, France
* andao@bu.edu

**Abstract:** In this paper, we proposed and numerically demonstrated a giant enhancement up to $10^8$ in both forward and backward propagation of the Second Harmonic Generation by combining the high-quality factor cavities of the Bound States in the Continuum and the excellent nonlinear optical crystal of lithium niobate. The enhancement factor is defined as the ratio of the second harmonic signal generated by the structure (lithium niobate membrane with Si grating) divided by the signal generated by lithium niobate membrane alone. Furthermore, a minimum interaction time of 350 ps is achieved despite the etching less lithium niobate membrane with a conversion efficiency of $4.77 \times 10^{-6}$. The origin of the enhancements is linked to the excitation of a Fano-like shape Symmetry-Protected Mode (SPM) that is revealed by Finite Difference Time Domain (FDTD) simulations. The proposed platform opens the way to a new generation of efficient integrated optical sources compatible with nano-photonic devices for classical and quantum applications.

### Introduction

In 1961, Peter Franken and colleagues used a seminal experiment to demonstrate for the first time the frequency doubling of light from a ruby laser beam focused into a quartz crystal [1]. Since then, nonlinear optical effects have played an essential role in frequency doubling and have been widely used in different applications, from classical and quantum light sources to interfacing devices operating at different frequencies in a quantum optical network given photon-based quantum information processing [2-4]. However, existing bulk materials suffer from weak optical nonlinearity, thus requiring high optical switching power or long interaction lengths. Therefore, nonlinear materials should be incorporated into a nanostructure with a high-quality factor to overcome the limitations. Bound states In the Continuum (BICs) are initially proposed in another field of wave physics, namely quantum mechanics, to achieve high-quality factors. They promise simultaneously prominent quality factors, compact devices, and local field enhancement, defined as the ratio of the total intensity divided by the intensity of the incident field [5-12]. However, BIC is a singularity that is often sensitive to fabrication imperfections. On the other hand, lithium niobate ($LiNbO_3$) is an excellent nonlinear optical crystal, owing to its high second-order susceptibility ($\chi^2$), large piezoelectricity, acousto-optic, and electro-optic coefficient features, as well as high refractive indices, but suffers from not trivial nanomanufacturing steps for applications requiring high-quality factor [13-19]. Thus, $LiNbO_3$ nanomanufacturing necessitates the use of costly equipment and a significant amount of effort. So far, proposed platforms for nonlinear effects are either based on waveguides or by nanomanufacturing the nonlinear materials, which limit the performance of the current reported second harmonic generation [20-24]. To fulfill this gap, metasurfaces and Mie resonance-based structures were recently proposed as alternative solutions [25-31]. However, the enhancements are limited. Here, we report a synergetic combination of the high Q-factor of the BIC and the excellent nonlinear optical crystal of $LiNbO_3$ to numerically demonstrate for the first time a giant enhancement of up to $10^8$ of the second harmonic generation in both forward and backward propagation. Moreover, the proposed platform does not require $LiNbO_3$ etching, thereby greatly simplifying the fabrication process. The platform consists of a mono-dimensional silicon grating (Si), separated by a low-index gap (air), deposited on a $LiNbO_3$ 900nm-thick lithium niobate membrane.

### Nano-structure design

To generate a giant second harmonic, we use the proposed slot waveguide unit cell. The advantage of using slot waveguides as unit cells is to integrate and overlap the high order susceptibility of the $LiNbO_3$ and the electric field of the BIC mode to generate a giant second harmonic. Fig. 1 shows the schematic of the proposed platform. The period of the unit cell (p) is 738 nm, the height of the unit cell (h) is 180 nm, the width (w) is 360 nm, the thickness of Si is 180 nm, and a X-cut or Y-cut self-suspended membrane of lithium niobate ($LiNbO_3$) with a thickness of 900 nm. Optimal parameters were chosen to enable SPMs to be excited in the Near Infrared (NIR) spectral range. Additionally, fabrication constraints were also considered, e.g., the aspect ratio AR = h/w must be less than 2 [32-35]. The structure is illuminated by a linearly polarized plane wave perpendicularly to the Si-grating.

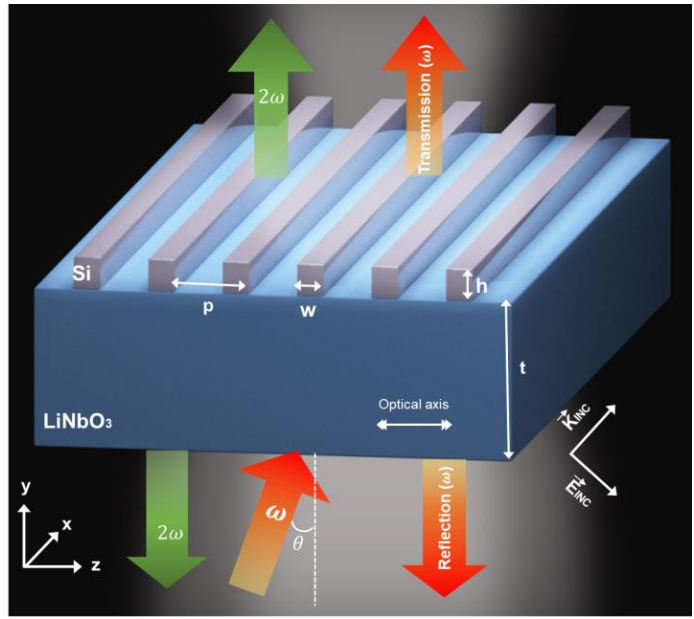

Fig. 1. Schematic of the proposed system made of Bound States in the continuum Cavities (BIC) on top of thin-film lithium niobate with a silica substrate. The grating consists of two high-index ridges (Si), separated by a narrow low-index gap (air). The incident wave is TM polarized with the electric field E parallel to the z-direction (for θ=0°), impinging the grating at oblique incidence. The geometrical parameters are period p, width, and height of the nano Si-grating, w and h respectively, and thickness of LiNbO$_3$ membrane t.

Due to the refractive index discontinuity, the proposed structure allows one of the propagating modes to confine its energy within the slot region [36-37]. The advantage of using this configuration is maximizing the confinement and interaction between the strong electric field supported by the slot waveguide and LiNbO$_3$ membrane. We performed numerical simulations using a homemade finite-difference time-domain (FDTD) code. We calculated the transmission spectrum as a function of the angle of incidence in the case of TM polarized light. Fig. 2 shows different modes that can be excited, as the SPMs (blue arrows) associated with the before-mentioned symmetry of the structure and LiNbO$_3$ membrane optical properties.

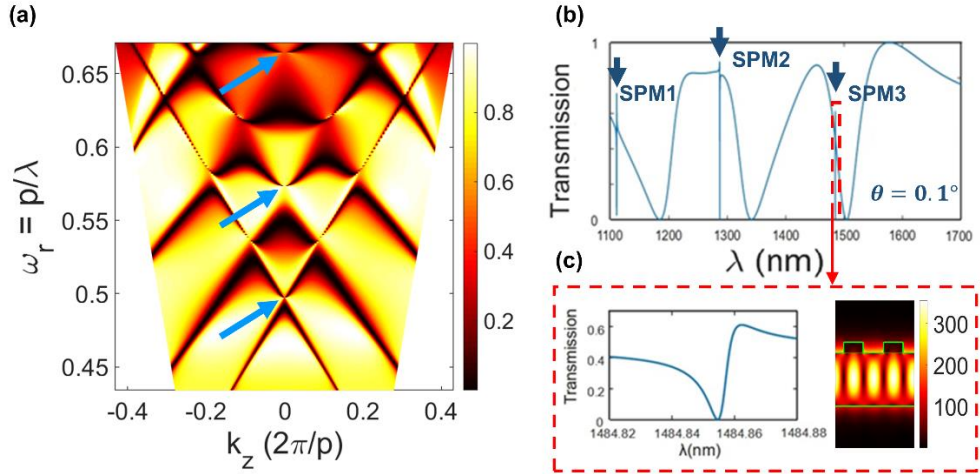

Fig. 2. (a) Dispersion curve in K-space (b) Transmission spectra of the linear signals in the case of a TM polarized (electric field perpendicular to the grating lines) plane wave impinging the structure at 0.1° incident angle. (c) Zoom of the third resonance peak in the spectrum. The inset represents the electric field amplitude of SPM3.

The SPMs are in the close vicinity of normal incidence, highlighting the modes' nature; the Q-factor decreases when the angle of incidence increases, and it tends slowly to infinity from its vanishing linewidth at normal incidence. Thus, illustrating the existence of the symmetric mode BIC. Hence, symmetry-breaking is necessary for the excitation of the modes with a high Q-factor. In this case, an oblique incidence 0.1° is selected. Fig. 2(b) shows the transmission spectrum calculated at this angle, where three high-quality-factor SPM's can be found, SPM1 at $\lambda_1$= 1110.52 nm, SPM2 at $\lambda_2$=1287.11 nm, and SPM3 at $\lambda_3$ = 1484.8545 nm. The corresponding quality factors for the modes are $Q_1$= 12400, $Q_2$ =12900, and $Q_3$ =221800, respectively. Specifically, we are interested in the highest Q-factor resonance of SPM3 to be used as a pump signal in order to generate an enhancement of the non-linear effect. A magnification of the SPM3 is presented in Fig. 2(b), showing the Fano-like spectral shape of the resonance. Such a high Q-factor in SPM3 leads to a better confinement of the light in the structure, with a constant mode volume. This could be quantified through the Purcell factor, which is higher as the quality factor increases and the mode volume decreases. The maximum electric field confinement in SPM3 was estimated to be larger than $\tau = \max(|E_{membrane+grating}|)/|E_{membrane}| = 350$ (see Fig. 3(g)). The electric intensity for both pump and SHG signals is the modulus squared of the electric field. Additionally, a Purcell factor of about $1.66 \times 10^5$ predicts a strong exaltation of non-linear phenomena in the LiNbO$_3$ membrane, Fig. 3 (g). It is worth noting the same geometrical parameters can also be used with a silica substrate to achieve similar efficiency. This highlights the robustness of our platform.

### FDTD method optimization

To take advantage of the largest non-linear coefficient of LiNbO$_3$, $d_{33}$, it is necessary to maximize the overlap between the localized SPM electric field distribution with the SHG. The second-order nonlinear polarization and the linear electric field relationship [14], are engineered, with an incident beam linearly polarized along the crystalline axis (Z) of LiNbO$_3$. This implies the use of a X-cut or Y-cut LiNbO$_3$ membrane so that the Z-axis can be perpendicular to the propagation direction. The relationship giving the nonlinear polarization of order two (PNL) for a Y-cut membrane can be written as:

$$\vec{P}_{NL}(2\omega) = \left(0, P_y(2\omega), P_z(2\omega)\right)$$

$$= \varepsilon_0 \begin{Bmatrix} 0 & 0 & 0 & 0 & d_{31} & -d_{22} \\ -d_{22} & d_{22} & 0 & d_{31} & 0 & 0 \\ d_{31} & d_{31} & d_{33} & 0 & 0 & 0 \end{Bmatrix} \begin{Bmatrix} 0 \\ E_y^2(\omega) \\ E_z^2(\omega) \\ 2E_y(\omega)E_z(\omega) \\ 0 \\ 0 \end{Bmatrix} \quad (1)$$

where the $d_{ij}$ tensor is the second-order susceptibility tensor of the nonlinear material, then for SHG we have the following equations:

$$P_y(2\omega) = \varepsilon_0\{d_{22}E_y^2(\omega) + 2d_{31}E_y(\omega)E_z(\omega)\}$$

(2)

$$P_z(2\omega) = \varepsilon_0\{d_{33}E_z^2(\omega) + 2d_{31}E_y^2(\omega)\}$$

The corresponding values for lithium niobate are: $d_{31}$= 5 x $10^{-12}$ pm/V, $d_{22}$= 3 x $10^{-12}$ pm/V an $d_{33}$ = 33 x $10^{-12}$ pm/V, respectively. After temporal and spatial discretization, Eq. (2) are integrated into a homemade FDTD code that combines two simultaneous electromagnetic simulations for ω and 2ω signals (Pump and SHG) involving 12 electromagnetic components. Eq. (1) is valid in the frequency domain meaning that it is necessary to calculate its Fourier transform before integrating it in the FDTD algorithm which operates in the temporal domain. Thus, the direct products of the pump field components, appearing in the second member of Eq. (1), turn into convolution products [38-39] (terms in E (ω) · E(ω) which leads in the temporal space to a convolution integral $\int E(t) \cdot E(t - t')dt$ ). To be numerically done, this convolution needs to store a huge amount of data. To circumvent this, some authors use the slowly varying envelope approximation which is not justified if one wishes to perform a broadband spectral study. Here, we have chosen the rigorously valid solution of a monochromatic calculation (wavelength by wavelength) which allows to transform the convolution product into a direct product.

### Results

More than 200 simulations were performed to calculate the SHG by varying the wavelength of the incident plane wave around the SPM one. The factor η evaluates the enhancement of the SHG:

$$\eta(\lambda) = \frac{I_{NL}^{grating}\left(\frac{\lambda}{2}\right)}{I_{NL}^{membrane}\left(\frac{\lambda}{2}\right)}$$

(3)

Where $I_{NL}^{grating}(\lambda/2)$ is the electric intensity of the total second harmonic signal generated by the membrane in presence of the grating and $I_{NL}^{membrane}(\lambda/2)$ corresponds to the total SH electric intensity generated by the membrane alone. Fig. 3 (a) illustrates the variation of the enhancement factor η as a function of the pump wavelength in backward and forward propagation. As expected, a huge value near $10^8$ is obtained at the SPM excitation with a larger Q-factor ($Q_{NL}$ = 330.000) than the pump signal; almost given by $Q_{NL} = \sqrt{2} \times Q_{NL}$. This is consistent with the fact that, to a first approximation, the intensity of the SH signal is proportional to the square of that of the pump, and if we model the resonance by a Gaussian (or Lorentzian) function, we easily obtain this relation between the two quality factors. Additionally, Fig. 3 (c)-(e), represent the SH fields and Fig. 3 (g)-(i), the pump electric intensity distribution inside the nanostructure at the three different SPMs. We can see a clear indication how the pump electric fields inside the structures are confined in each of the SPMs resonances, as well as an evidence of the enhancement at SPM3 (SHG enhancement) Fig. 3 (i).

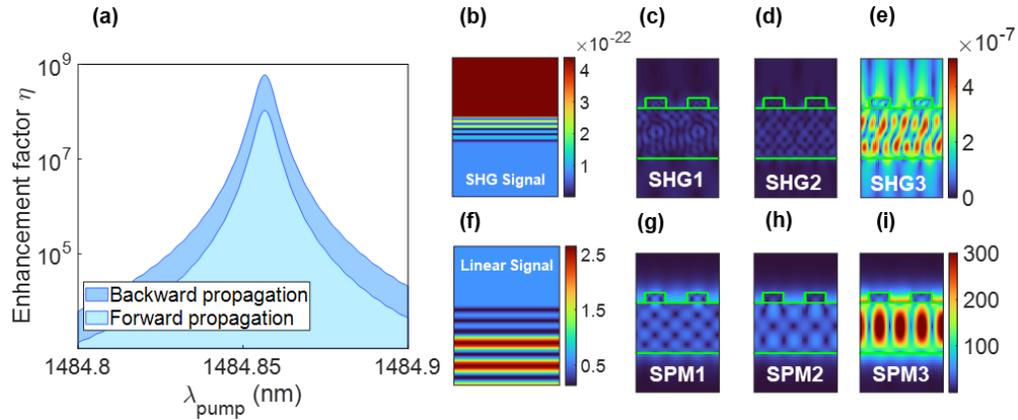

Fig. 3. (a) SHG Enhancement factor η spectrum at backward and forward propagation at SPM3. The enhancement factor is defined as the ratio of the SH signal generated by the structure (lithium niobate membrane with Si grating) divided by the signal generated lithium niobate membrane alone. (b) The SH signal in unstructured lithium niobate membrane alone and (c)-(e) electric field amplitude distributions in two periods structure for the three SPMs as indicated in Fig. 2(b). (f) Electric field amplitude of the Pump signal in the case of the lithium membrane alone. (g)-(i) Spatial

distributions of the electric field amplitude of the Pump at the three SPM's resonances. (c) and (g) SPM1, wavelength $\lambda_1 = 1110.52$ nm. (d) and (h) SPM2, wavelength $\lambda_2 = 1287.11$ nm. (e) and (i) SPM3, wavelength $\lambda_3 = 1484.8545$ nm.

### Fabrication imperfections analysis

We perform the fabrication imperfection analysis variating $\pm\Delta w$, $\pm\Delta p$, $\pm\Delta t$, and $\pm\Delta h$ to evidence the possible shifts of the SPMs. The models are defined with the parameters $\pm\Delta p$, $\pm\Delta t$, and $\pm\Delta h$ equal to 10 nm and with the parameter $\pm\Delta w$ equal to 8 nm. We focus on the resonance of interest SPM3, where we can notice that in Fig. 4 (a)-(c), the SPM3 location in the transmission spectrum is constant after the different variations. Therefore, after variations in the parameters w, t, and h, the SPM3 changes are unperceived. However, the structure shows increased sensitivity to the parameter p. As we can see in Fig. 4 (d), the period could influence SPM3 resonance shifts. This result is expected and easy to avoid using standard fabrication process.

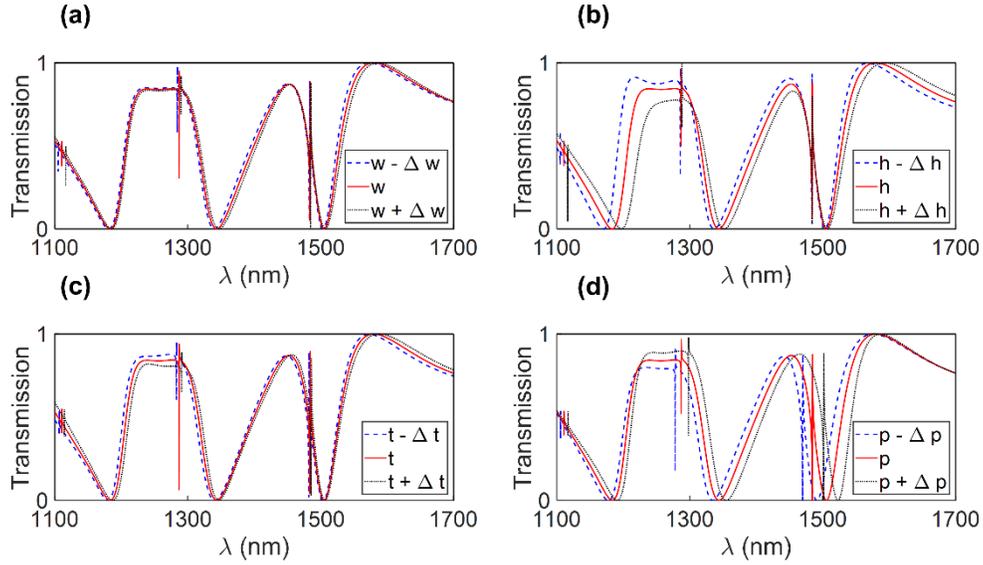

Fig. 4. (a)-(d) Transmission spectra for imperfection analysis of parameter w, h, t, p. Fabrication imperfection analysis variating the dimension at $\pm\Delta t$, $\pm\Delta p$ and $\pm\Delta h = 10$ nm. $\pm\Delta w = 8$ nm

To present a real application concept, it is necessary to pass from an ideal case, as an infinite grating nanostructure, to a real one. To analyze the best dimensions to achieve the desired nonlinear effects and the physical phenomena (SPMs resonances), an examination of the optical response (pump signal only) of a finite structure composed of Np periods (from Np = 100 to 1000) is necessary. Therefore, the real size of the structure then will vary approximately between 75 µm and 750 µm. Thanks to its low aspect ratio (w/h = 0.5) and its duty cycle (w/p=0.5), the structure can be appropriately fabricated without real difficulties using standard fabrication process.

We perform a rigorous analysis where the number of periods is varied, Fig.5 (Left); we have considered an illumination by a Gaussian beam, where its beam-waist is adapted to the structural dimension. For example, for a structure of Np = 1000 periods with a total dimension of $D = 1000 \times 738$ nm $\approx 738$ µm, the beam-waist is adjusted to be equal to $D/2 = 370$ µm meaning that the incident electric field at the edges of the structure is equal to $1/e$ its value at the center. In the case of these finite structures, the transmission is defined by the ratio of the total transmitted power to the total incident one. In another word, it is the ratio of the Poynting vector flux through a surface parallel to the membrane calculated from the transmission side to the same quantity calculated from the incident side of the structure.

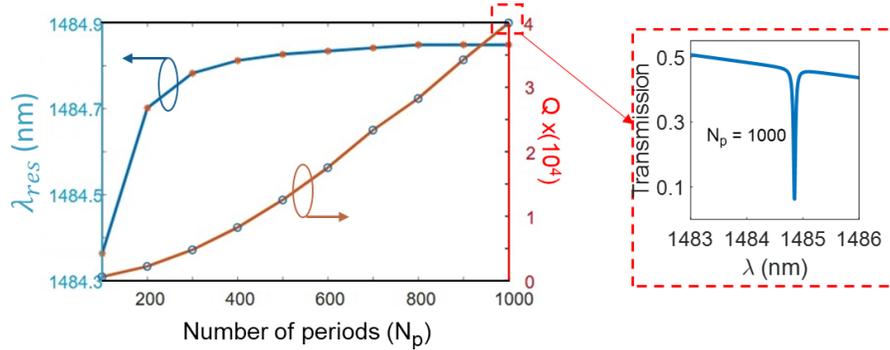

Fig. 5. (Left) Influence of the number of periods $N_p$ on the Si-grating at the SPM3 resonance wavelength (blue line) and its $Q-$factor also as a function of the period number $N_p$ (orange line). (Right) Example of transmission spectrum for a finite structure composed of $N_p = 1000$ periods.

Fig. 5, shows the variations of both $\lambda_{res}$ and Q-factor as a function of the grating period number. It is easy to see that the quality factor of the resonance varies strongly with $N_p$ reaching a value of $Q \approx 40000$ for $N_p = 1000$ while the resonance wavelength tends asymptotically, but rapidly, to the one of the infinite structure. An example of transmission spectrum is given in Fig. 5 (right) for $N_p = 1000$ periods. In this case, a resonance dip occurs at $\lambda_{res}$=1484.8 nm. This result demonstrates the need to fabricate very large structures in order to recover the resonance properties of the infinite grating.

**Conclusion**

In summary, we have proposed and numerically demonstrated a giant enhancement of the second harmonic generation up to $\mathbf{2 \times 10^8}$ in both forward and backward propagation, evidencing a small screening effect of the silicon grating presence. Furthermore, the time interaction between light and the structure to generate such enhancement is inversely proportional to the Q-factor of the resonance, giving rise here to a minimum interaction time of 350 ps. Because of the small duty cycle w/p=0.5 and aspect ratio h/w=0.5, proposed platform can be easily realized using a standard fabrication process. The proposed platform opens the way to a new generation of efficient integrated optical sources compatible with nano-photonic devices for classical and quantum applications.